\documentclass[aps,prl,twocolumn,showpacs,superscriptaddress,groupedaddress,nofootinbib]{revtex4}  
\usepackage{graphicx}  
\usepackage{dcolumn}   
\usepackage{bm}        
\usepackage{amssymb}   
\usepackage{amsmath,amsthm,latexsym,amssymb,amsfonts,epsfig}
\usepackage{float}
\usepackage{adjustbox}
\usepackage{multirow}
\usepackage{tabularx}
\usepackage[latin9]{inputenc}
\usepackage[usenames,dvipsnames]{xcolor}
\newcommand{\blue}{\textcolor{blue}} 

\newcommand{\be}{\begin{equation}}
\newcommand{\ee}{\end{equation}}
\newcommand{\bea}{\begin{eqnarray}}
\newcommand{\eea}{\end{eqnarray}}

\begin{document}

\widetext


\title{Emergent dark energy from unparticles}
\author{Micha{\l} Artymowski}
\affiliation{Physics Department, Ariel University, Ariel 40700, Israel}
\affiliation{Cardinal Stefan Wyszynski University, College of Science, Department of Mathematics and Natural Sciences, Dewajtis 5, 01-815 Warsaw, Poland}
\author{Ido Ben-Dayan}
\affiliation{Physics Department, Ariel University, Ariel 40700, Israel}
\author{Utkarsh Kumar}
\affiliation{Physics Department, Ariel University, Ariel 40700, Israel}
\date{\today}

\begin{abstract}
A limiting temperature of a species can cause the Universe to asymptote to it yielding a de-Sitter (dS) phase due to macroscopic emergent behavior. The limiting temperature is generic for theories slightly shifted from their conformal point. We demonstrate such behavior in the example of unparticles/Banks-Zaks theory.
The unparticles behave like radiation at high energies reducing the Hubble tension, and a cosmological constant (CC) at low energies
yielding a model that follows closely $\Lambda$CDM model but due to collective phenomenon. It is technically natural and avoids the no-dS conjecture. The model is free of the coincidence and initial conditions problems, of scalar fields and of modified gravity. 
\end{abstract}

\maketitle

\section{Introduction} \label{sec:intro}

The cosmological data from the CMB \cite{Aghanim:2018eyx} as well as the discovery of the acceleration of the Universe \cite{Riess:1998cb,Perlmutter:1998np} strongly suggest that the Universe is partially filled with Dark Energy (DE) \cite{Bamba:2012cp}, which currently constitutes around $70\%$ of the energy density of the Universe. The simplest model that explains these measurements assumes that DE is a CC with the energy density of order of $\rho_{DE}\sim 10^{-119}M_p^4$, where $M_p \simeq 2.435 \times 10^{18}GeV$ is the reduced Planck mass. 
A popular alternative is that DE originates from some dynamical degree of freedom, due to a modification of gravity or additional scalar fields \cite{Copeland:2006wr}. The present-day value of $\rho_{DE}$ is a source of fine tuning in several ways. Assuming DE is a true CC (or a fluid with $w\simeq -1$ throughout the whole evolution of the Universe up to today) one obtains $\rho_{DE} \lll M^4$, where $M$ could be taken as any fundamental scale of known physics, such as $M_p\simeq 10^{18}GeV$, $M_{EW}\sim 10^2GeV$ or $M_{QCD}\sim 0.3 GeV$ \cite{Carroll:2000fy}. The huge hierarchy between $\rho_{DE}$ and other energy densities in the early Universe is often labeled as a problem of initial value of DE \cite{steinhardt,Dinverno}. Another issue is the so-called coincidence problem \cite{Zlatev:1998tr,Velten:2014nra}. Since most of the evolution of the Universe happened in eras of radiation or dust domination, it is a rather big coincidence that "nowadays" energy densities of dust and DE are of the same order of magnitude. A less concerning problem is the Hubble tension, suggesting $\gtrsim4\sigma$ discrepancy between the value of the Hubble parameter measured by late Universe observations $(z\lesssim 1)$ compared to early Universe ones $(z\gg1)$ \cite{Verde:2019ivm}. Finally, the no-dS conjecture stipulates that true dS vacua or long term dS like phases such as Inflation or Dark Energy are problematic, if not completely forbidden according to our knowledge of scalar fields in quantum gravity theories \cite{Ooguri:2018wrx,Ben-Dayan:2018mhe,Artymowski:2019vfy}. We show that considering a broken conformal field theory close to its conformal point may be a solution to all these problems. Furthermore, this behavior is generic for such theories. Finally, the resolution of these problems is not due to a specific fundamental degree of freedom, but due to the collective behavior of the theory.

One of the underlying assumptions of the aforementioned problems of cosmology is the use of perfect fluids with $p_i=w_i \rho_i$, where $w_i$, the equation of state parameter for each species is approximately constant and specifically, there exists some inflaton/quintessence field, such that $w_{DE}\simeq -1$ for a long enough duration. This is achieved, for instance, by tuning the potential of the inflaton/quintessence to be flat. As such, at some point in time $\dot{H}=-\frac{1}{2}\sum_i(1+w_i)\rho_i\rightarrow 0$ marking the onset of the dS phase. A path less traveled is discarding the scalar field, and analyzing the macroscopic behavior of a sector resulting in, $w\neq const.$ While this is not the case considering standard matter and radiation, it is the generic situation of broken conformal field theories (CFT) close to a conformal fixed point. In such a case the trace of the energy momentum tensor $\theta^{\mu}_{\mu}=\rho_u-3p_u\propto \beta(g)=c T^{x}$, where $x$ is some anomalous scaling, and $c$ is dimensionful \cite{Grzadkowski:2008xi,Artymowski:2019cdg}\footnote{A more general situation can be if we simply demand $\theta^{\mu}_{\mu}=\rho-3p=f(T)$ for some function $f$, with proper dimensions.}. 
This result is based solely on dimensional analysis, and is valid to all CFTs. Considering perfect fluids on top of the broken CFT without additional couplings, results in $\dot{H}=-\frac{1}{2}\left(\sum_i(1+w_i)\rho_i+\frac{4\rho_u-c T^{x}}{3}\right)$, and we can reach the limit $\dot{H}\rightarrow 0$ at some temperature $T_c$ of $\rho_u,p_u$. As a result, the temperature of this species can approach a constant at some temperature, and the species behaves as a CC, while the rest of the species in the Universe continue to cool due to its expansion.

 We shall consider a specific example in the framework of the Banks-Zaks theory \cite{Banks:1981nn,Georgi:2007ek}. At high temperatures (i.e. for $T\gg \Lambda_\mathcal{U}$, where $\Lambda_\mathcal{U}$ is a cut-off scale of the theory) one finds the Universe with the standard model (SM) sector coupled to Banks-Zaks particles with energy density $\rho=\sigma_{BZ}T^4$. The coupling gives the anomalous dimension to BZ. This radiation like behavior, adds relativistic species and increases $N_{eff}$ partially relieving the Hubble tension. Below the scale $\Lambda_\mathcal{U}$ the BZ sector decouples from SM and BZ sector can be described as an unparticle ``stuff'' \cite{Georgi:2007ek,Grzadkowski:2008xi} with anomalous scaling. 
Under certain conditions, using the thermal average of the theory \cite{Grzadkowski:2008xi}, the BZ sector will asymptote to a limiting temperature \cite{Artymowski:2019cdg}, yielding a valid DE behavior \footnote{
 In \cite{Dai:2009mq} the authors considered scalar unparticles with a mass as a function of  scaling dimension of unparticles. Unparticles  have been studied in framework of general relativity and loop quantum cosmology \cite{Chen:2009ui,Jamil:2011iu} where authors discuss the stability of unparticles interacting with standard radiation.}. The model resolves the initial conditions problem because it does not need a small CC, or a small initial $\rho_{DE}$ since, its energy density at early times is similar to radiation. Hence, the model has a built-in tracker mechanism. The solution is technically natural, since taking a small parameter to zero reproduces a conformal symmetry in the unparticle sector. 
The absence of a fundamental scalar field makes the theory immune to the no-dS conjecture. Finally, since the present acceleration is given by the collective behavior of the BZ sector, it could be that the true CC is zero, possibly solvable by some symmetry argument, such as conformal symmetry \cite{Ben-Dayan:2015nva}, reverting us back to the "Old CC problem" of making the CC vanish.

\section{Unparticles in FLRW Universe} \label{sec:unparticles}

In this work we consider the unparticles as a possible candidate for DE. A successful DE model must satisfy several conditions, namely:
\begin{itemize}
\item The present-day value of DE energy density is $\rho_{DE} \simeq 1.7 \times 10^{-119} M_p^4$.
\item The energy density of DE must be subdominant at the BBN / CMB scale, in order to satisfy the BBN  and $N_{eff}$ constraints i.e. $\frac{\rho_{DE}}{\rho_{r}} \leq 0.086$ \cite{Cooke:2013cba,Artymowski:2017pua,Ben-Dayan:2019gll} 
at $95\%$ confidence level.
\item Between eras of radiation and DE domination one must have an era of dust domination, which is essential to the growth of the large scale structure of the Universe. 
\item The equation of state of DE today defined as $w_{DE}\equiv\frac{p_{DE}}{\rho_{DE}}$ where $\rho_{DE}$ and $p_{DE}$ are energy and pressure of DE respectively, must lie within $ -1.14 < w_{DE}< -0.94 $ \cite{Aghanim:2018eyx}. In our model, imposing the previous constraints automatically fulfills this requirement.
\end{itemize}

Consider the flat universe filled with unparticles and the \blue{perfect} fluids of matter and radiation.
In such case, the Friedmann equations are
\begin{eqnarray}
3 H^{2} & = & \rho = \rho_u + \rho_r + \rho_m  \label{eq:frw1} \, ,  \\
\dot{H} & = & -\frac{1}{2} \left( \rho + p \right) = -\frac{1}{2}\left(\rho_u + p_u + \rho_m + \frac{4}{3}\rho_r\right)\, ,
\end{eqnarray}
where $\rho_m \propto a^{-3}$ and $\rho_r\propto a^{-4}$ are energy densities of dust and radiation respectively.
Following \cite{Grzadkowski:2008xi,Artymowski:2019cdg}, the energy density and pressure of unparticles are
\begin{eqnarray}
 \rho_u & \simeq & \sigma T^4 + B \,T^{4 + \delta} = \sigma\, T_c^4\, y^4 \left(1-\frac{4 (\delta +3) y^{\delta }}{3 (\delta
   +4)}\right)\, , \label{eq:rhou}\\
 p_u &\simeq& \frac{\sigma}{3}  T^4 + \frac{B}{\delta + 3} T^{4 + \delta} = \frac{\sigma}{3} \, T_c^4 \, y^4 \left(1-\frac{4 y^{\delta }}{\delta
   +4}\right) ,\label{eq:pu}
\end{eqnarray}
where $\delta$ is associated with the anomalous dimension accordingly: $\delta=a+\gamma,\, \beta(g) = a(g-g_\star)$ is the beta function, $a>0$, $g_\star$ is the value of the coupling at the fixed point, and $\gamma$
is the anomalous dimension of the operator. $\delta<0$ corresponds to $\gamma<-a<0$. $\sigma$ is the number of the degrees of freedom in the sector, $y=T/T_c$ is a dimensionless temperature and $T_c=\left[\frac{4(\delta+3)}{3(\delta+4)}\left(-\frac{\sigma}{B} \right)\right]^{\frac{1}{\delta}}$, is the temperature at which $p_u=-\rho_u$. Let us stress that Eqs. (\ref{eq:rhou},\ref{eq:pu}) are merely approximate values, which are valid below the cut-off scale \cite{Tuominen:2012qu}. Above the cut-off we should recover radiation-like behavior. This naturally happens in the case of negative $\delta$ in our approximation, since for $T \gg \Lambda_U$ one finds $p_u/T^4 \to const$ and $(\rho_u-3p_u)/T^4 \to 0$. Hence, if there are significant deviations from our approximation, they will have to cancel one another quite accurately to recover the radiation-like behavior. So in the case of negative $\delta$ our approximation should remain perturbative up to the cut-off, above which the anomalous term should vanish, and we are left with the standard $\rho_u\sim T^4$ radiation-like behavior \cite{Tuominen:2012qu}.

\begin{figure}[]
\centering
 \begin{adjustbox}{center}
\includegraphics[width=4.2cm, height=3.8cm]{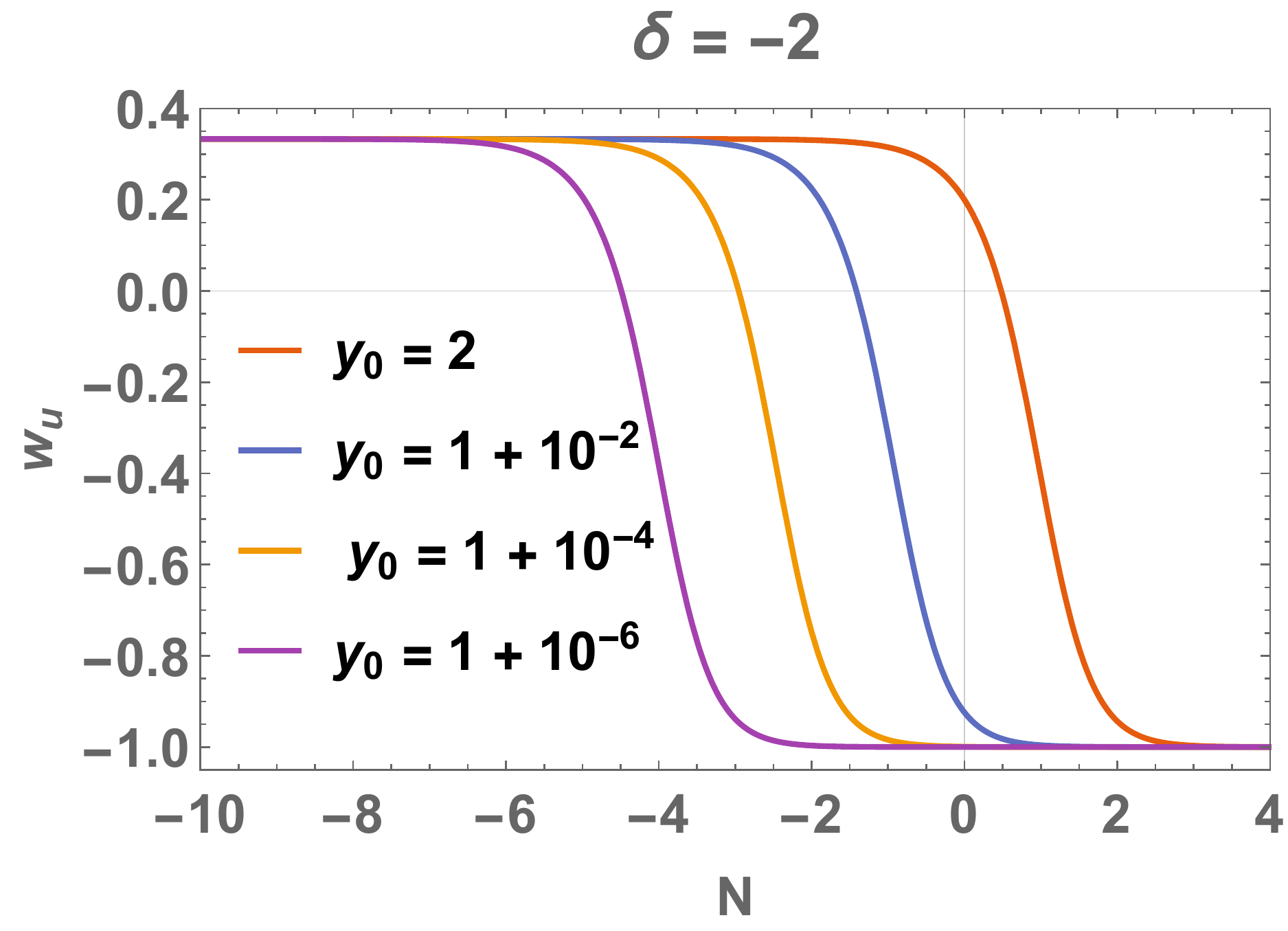}
\includegraphics[width=4.2cm, height=3.8cm]{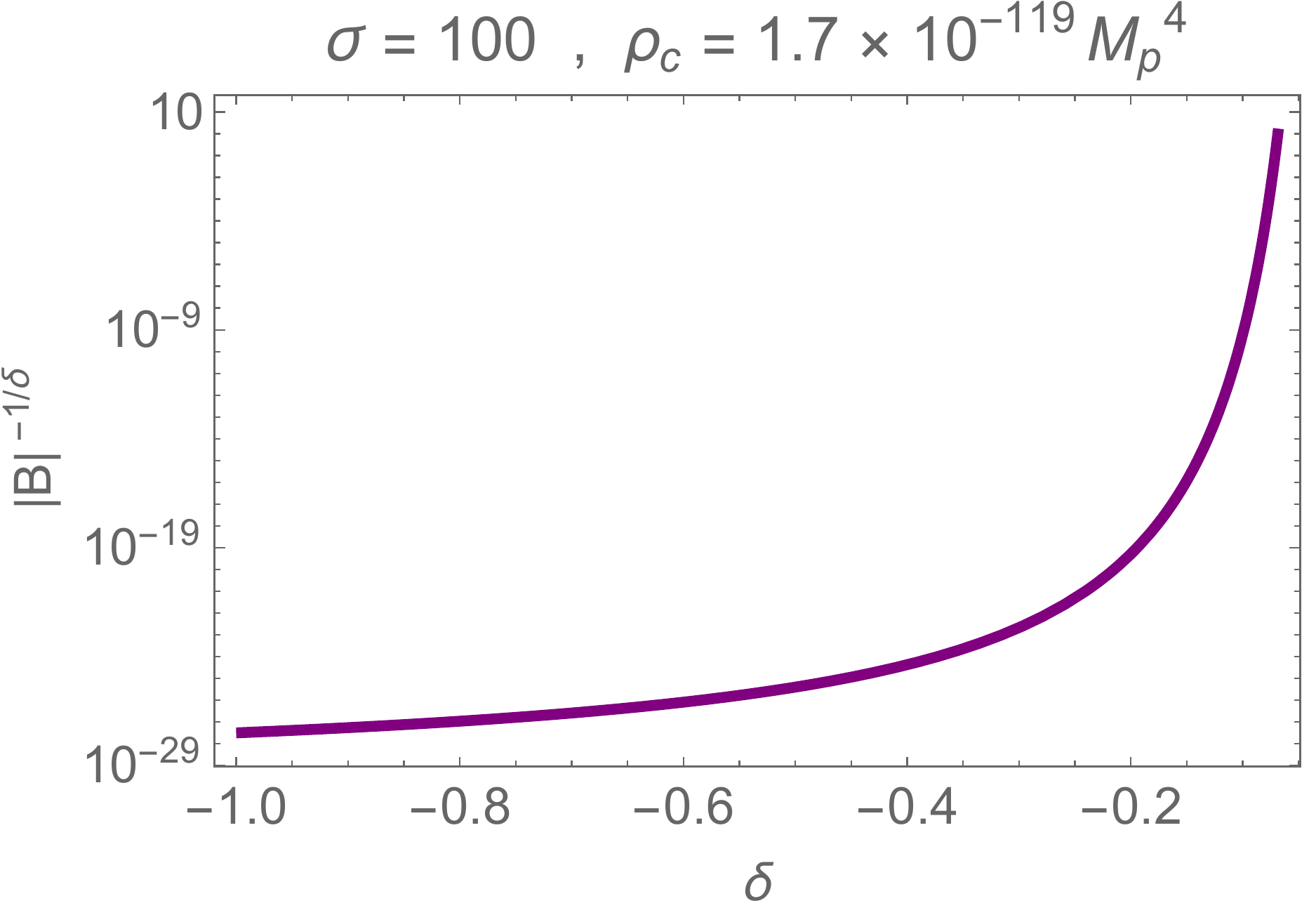}
\end{adjustbox}
\caption{\it Left panel: The equation of state of unparticles $w_u$ as a function of e-folds, $N$, for $\delta=-2$ and different initial conditions. The unparticles have reached a CC behavior $y_0\simeq 1$.  \it Right panel: 
 The energy scale $B^{-1/\delta}$ which has a dimension of mass with $M_{p}=1$ vs. $\delta$ for $\sigma=100$. The $B(\delta)$ is obtained by matching $\rho_\infty$ to the energy density of a present day CC in $\Lambda$CDM. Note that $\delta  \sim -0.068$ results in $B^{-1/\delta}\simeq M_p$.
} 
\label{fig:tracker}
\end{figure}

From the continuity equation of unparticles $\dot{\rho}_u=-3H(\rho_u+p_u)$, we find that at $T_c$ the energy density of unparticles will become constant. Additionally, we use it to solve for the scale factor,
\begin{equation}
a(y) = \frac{y_0}{y} \left(\frac{1-y_0^{\delta }}{1-y^{\delta }}\right)^{\frac{1}{3}} \, , \label{eq:aofy}
\end{equation}
where $y_0$ is the present-day value of $y$, $a(y_0) = 1$, and $y=1$ corresponds to future infinity, while in the past $y\gg1$. The form of $a(y)$ depends only on $\delta$ and $T_c$, which are parameters of unparticles. The form of \eqref{eq:aofy} does not change if one considers more fluids filling the Universe. So unparticles act as the "clock" of the Universe. Positive $\rho_u$ and $T_c$ require $-3\leq \delta \leq 0$.\footnote{Let us note, that both $\delta=-3$ and $\delta=0$ are well defined limits, see \cite{Artymowski:2019cdg}. For $\delta=0$ one has $\rho_u=\sigma T^4+3 A \, T^4 \ln T,\, p_u=(\sigma-A)/3 \, T^4+A \, T^4 \ln T$. It preserves the most important feature of the DE candidate, namely for $T=T_c$ one finds $\dot{H} = 0$, with $T_c=e^{\frac{A-4\sigma}{12A}}$. For any finite $\delta$ there is a relation $A = \frac{\delta B}{3+\delta}$. Eqs. \eqref{eq:small_delta1},\eqref{eq:small_delta2} are exact in the $\delta=0$ case.}
In this parameter range unparticles always fulfill the Null Energy Condition, $w\geq-1$. If unparticles are in thermal equilibrium with the SM at early times their temperature should be similar to radiation $T\sim T_r$. In the $y\gg1$ regime one finds $\rho_u \propto y^4 \propto T^4$, $a\propto 1/T$ and $\rho_u/\rho_r \to const$. As a result, in the early Universe unparticles behave like ordinary radiation\footnote{This is expected, since for $T>\Lambda_\mathcal{U}$ one should recover BZ theory with $\rho_{BZ}=\sigma_{BZ}T^4$.} and the Universe evolves like the standard $\Lambda$CDM, but with some additional relativistic degrees of freedom, $N_{eff}$. As the Universe cools the unparticles decouple and have a different temperature. At late times, $\dot{\rho}_u   \xrightarrow{y\xrightarrow{}1} 0$, and unparticles asymptote to a CC. This feature is most easily demonstrated in the $|\delta|\ll1$ limit, which results in an equation of state and scale factor independent of $\delta$ at first order:
\begin{eqnarray} \label{eq:small_delta1}
w_u\equiv\frac{p_u}{\rho_u}\simeq \frac{1}{3}\frac{\ln y-1/4}{\ln y+1/12} \, ,\\
a(y)\simeq\frac{y_0\ln^{1/3}y_0}{y\ln^{1/3}y} \, . \label{eq:small_delta2}
\end{eqnarray}
Notice that at $y\gg 1$, $w_u\simeq 1/3$ while at $y\simeq 1$, $w_u\simeq -1$. Therefore, unparticles have a built-in "tracker mechanism" which is a crucial difference between unparticles and scalar field models with tracker solutions, for which the equation of state of the field becomes close to $-1$ during the radiation domination era (see Ref \cite{Huterer:2017buf} for details). The other regime that can be investigated analytically
is the late time, $y\gtrsim 1$ : 
\begin{eqnarray}
y(N) &\simeq& 1 + e^{-3N}\left( y_0 -1 \right)\, , \label{eq:approxy(N)} \\
 \rho_{u}(N) &\simeq& -\frac{\delta  \sigma  T_c^4}{3 (\delta +4)} \left( 1 + e^{-3 N} 4 (\delta +4)  \left(y_0-1\right)\right) \, , \label{eq:approxyrho(N)} \\
   w_{u} & \simeq & -1 + e^{-3 N}\, 4 (\delta +4) \left(y_0-1\right) \, , \label{eq:approxwu}
\end{eqnarray}
where $N\equiv \log(a)<0$ are the so called e-folds, $N=0$ today. From Eqs. \eqref{eq:approxy(N)} and \eqref{eq:approxyrho(N)} one can see that the temperature and energy density of unparticles decreases with $N$ (and therefore with time) to obtain a constant values $T\simeq T_c$ and $\rho \simeq \rho_\infty \equiv -\frac{\delta \, \sigma}{3(\delta+4)}T_c^4$. As $N$ increases radiation and dust become negligible and unparticles dominate at late times.\footnote{Following the procedure described in \cite{Nojiri:2005sx,Copeland:2006wr} we find that our model is also free from any type of future singularity.} Notice that the approach of $\rho_u, w_u$ to a constant is exponential in e-folds, therefore the deviations from $\Lambda$CDM are expected to be very small at low redshift.
An example of the behavior of the equation of state parameter of unparticles $w_u$ as a function of $N$ for $\delta=-2$ is given in left panel of Fig \ref{fig:tracker}. At early times $w_u=1/3$, while today, $w_u=-1$, and $T_{u}\simeq T_c$.

\begin{figure}[]
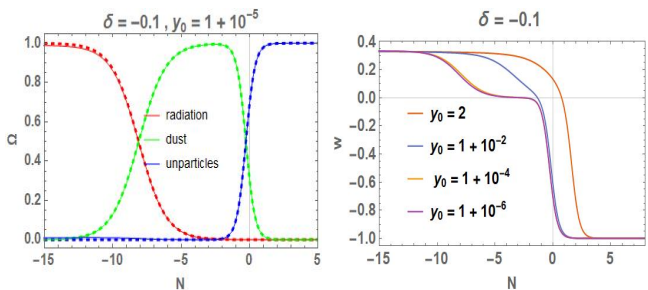

\centering
\begin{adjustbox}{center}
\includegraphics[width=4.2cm, height=3.8cm]{energydensity.pdf} 
\includegraphics[width=4.2cm, height=3.8cm]{wtot.pdf}
\end{adjustbox}
\caption{\it Left panel: Comparison of the evolution of density parameters $ \Omega(N)$ for each fluid between $\Lambda CDM $(dashed) and unparticles (solid) model. The dashed blue curve is the relative density of the CC in the $\Lambda$CDM model. Right panel: Evolution of the total equation of state $w(N)$. Both $\Omega$ and $w$ weakly depend on $\delta$. The universe evolves from the radiation-dominated phase to matter domination era followed by DE domination (i.e. the unparticles domination) with $w\simeq -1$. If $y_0-1\gtrsim 10^{-2}$ the matter-radiation equality and/or the total $w$ do not fit the data. }
\label{fig:analytic_d-3}
\end{figure}

Let us define the usual density parameters,
\begin{equation}
\Omega_m\equiv\frac{\rho_m}{\rho} \, , \qquad \Omega_r\equiv\frac{\rho_r}{\rho} \, , \qquad 
\Omega_u\equiv\frac{\rho_u}{\rho} \, .
\end{equation}
According to the Planck data, present-day values of these parameters are equal to $\Omega_m^0 = 0.3089 $, $\Omega_r^0 = 8.97 \times 10^{-5}$ and $\Omega_u^0 = 0.6911$ \cite{Aghanim:2018eyx}, assuming unparticles are responsible for the present day acceleration. This fitting constrains $B$ for a given $\delta$ and $\sigma$ as shown in right panel of Fig. \ref{fig:tracker}.
Notice that this energy scale could be in a huge span of energies, $10^{-30} M_p<B^{-1/\delta}<M_p$, for $\sigma=100$. 
Finally, in Fig. \ref{fig:analytic_d-3} we show an example of the evolution of energy densities and the total equation of state. The results are not very sensitive to $\delta$ but require $y_0-1<10^{-3}$ for a long enough dust domination era. 

\section{Early Universe evolution and BBN/CMB constraints}\label{sec:BBN}

\begin{figure}[]
\centering
\begin{adjustbox}{center}
\includegraphics[width=4.2cm, height=3.0cm]{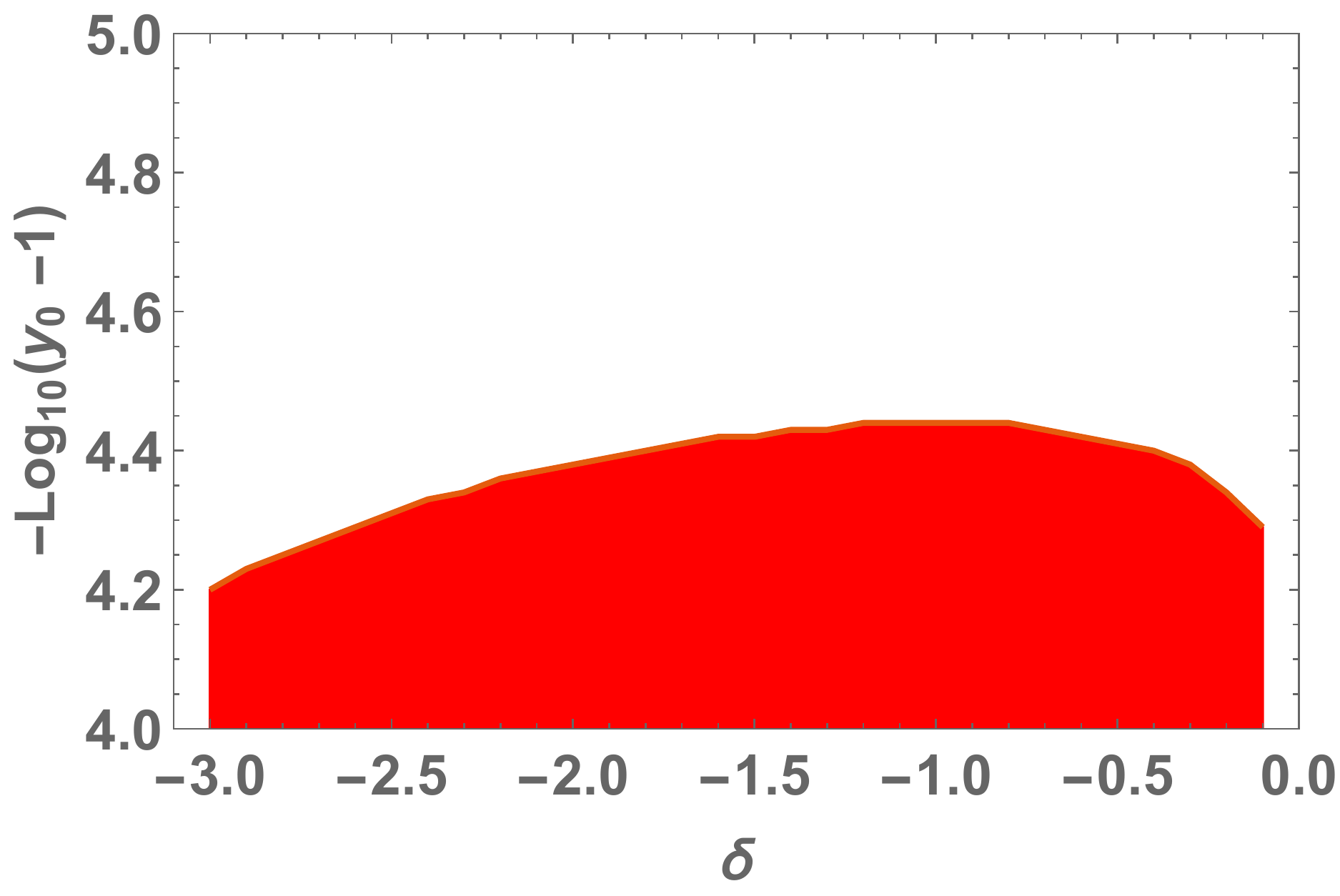} 
\vspace{0.5cm}
\includegraphics[width=4.2cm, height=3.0cm]{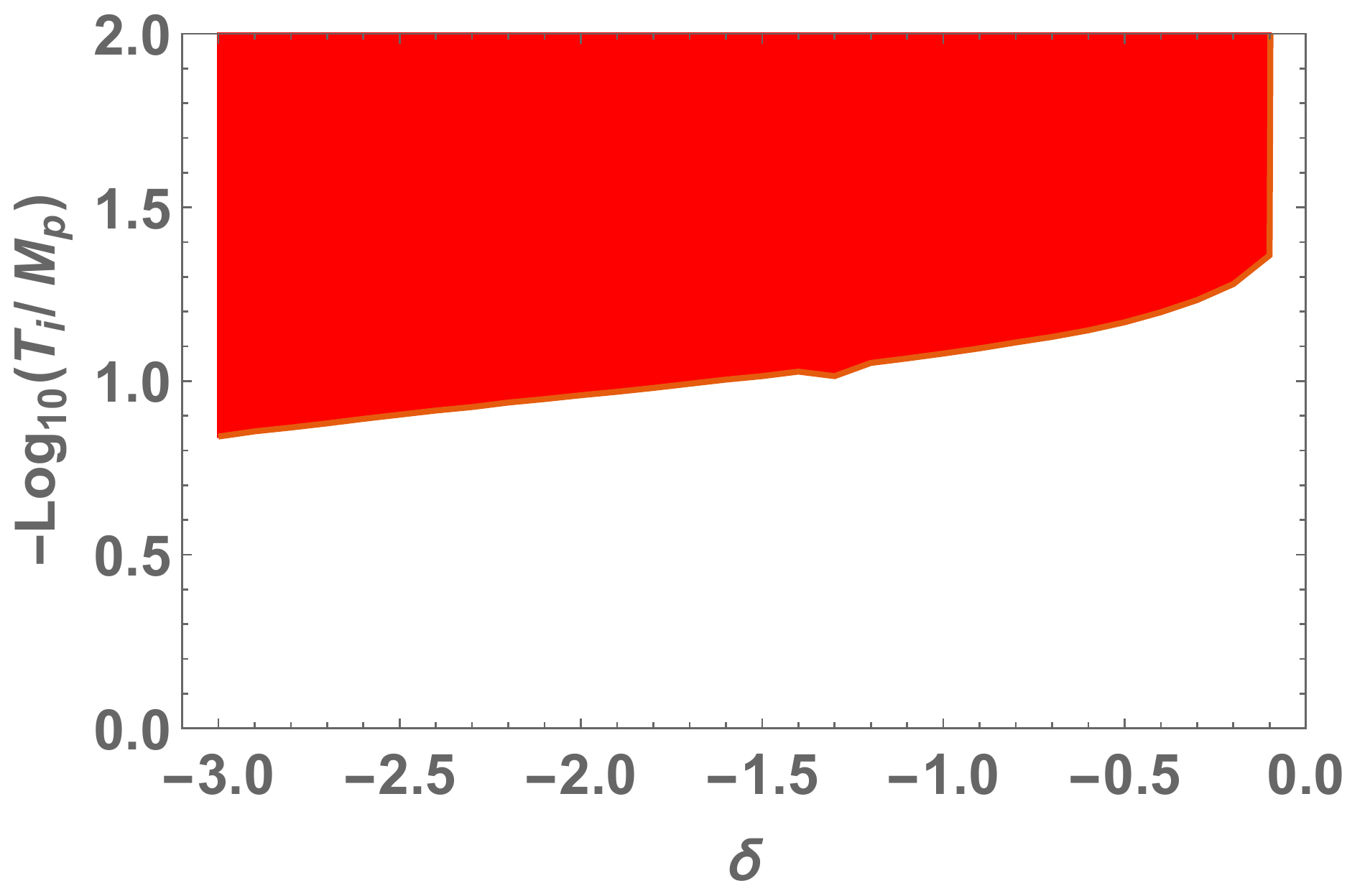} 
\end{adjustbox}
\caption{\it Left panel: $-\log_{10}(y_0-1)$ vs. $\delta$. White regions of parameter space are consistent CMB constraints. Right panel: BBN constraints on the $(\delta,T_i)$ parameter space, where $T_i$ is a value of $T_u$ for $\rho_r\simeq \rho \simeq M_p^4$. Note that any value $T_u < 0.1 M_p$ gives a viable $\rho_u/\rho_r$ at BBN/CMB.} 
\label{fig:BBN}
\end{figure}

As mentioned in the previous section, in the $y\gg1$ limit one finds $y^\delta\ll1$ and therefore $\rho_u \propto y^4\propto a^{-4}$. Thus, in the early Universe one should expect $\rho_u/\rho_r$ to be constant. Indeed, knowing $\Omega_{u0}$, and $\Omega_{r0}$ one finds in the $y \gg 1$ limit and $y_0\simeq 1$:

\begin{equation}
\frac{\rho_u}{\rho_r}\simeq \frac{\Omega_{u0}}{\Omega_{r0}} 3(\delta+4)(-\delta)^{1/3}(y_0-1)^{4/3}.
\end{equation}
One can see that the ratio between $\rho_u$ and $\rho_r$ depends only on values of $\delta$ and $y_0$, as $\Omega_{u0}$ and $\Omega_{r0}$ are fixed by the data. On the other hand, the big bang nucleosynthesis (BBN) and cosmic microwave background (CMB) constrain the allowed number of relativistic degrees of freedom at BBN and decoupling \cite{Cooke:2013cba,Artymowski:2017pua,Ben-Dayan:2019gll,Kneller:2002zh} 
\begin{equation}
\left.\frac{\rho_u}{\rho_r}\right|_{BBN} \leq \frac{7}{8}\left(4/11\right)^{4/3}2\Delta N_{eff}\simeq 0.086 \, ,
\end{equation}
at the $95\%$ confidence level, where $\Delta N_{eff}=3.28-3.046=0.19$ \cite{Ben-Dayan:2019gll}.
Thus, from the BBN/CMB data one can constrain the $(\delta,y_0)$ parameter space, which is presented in the left panel of Fig. \ref{fig:BBN}. For $y_0-1\lesssim 10^{-5}$ one satisfies the BBN/CMB constraints for all $\delta$. 
Taken at face value, the addition of unparticles ameliorates the Hubble tension to $\simeq2-3\sigma$, as it pushes the CMB derived Hubble parameter towards $H_0\simeq 70\, km/Mpc/sec$,\,\cite{Simha:2008mt}. Of course, a full likelihood analysis should be performed for correct inference.

The BBN constraint can also be translated into allowed range of temperatures in the high energy limit. One can take the allowed range of $y_0$ and evolve it back to high energies, which gives the upper bound on $T_u$ presented in the right panel of Fig. \ref{fig:BBN} taken at the Planck scale, i.e. for $\rho \simeq \rho_r = M_p^4$. Note that any value $T_u < 0.1 M_p$ gives $\left.\rho_u/\rho_r\right|_{BBN}$ consistent with the data.

Of specific interest is the $\delta=0$ limit. In such a case, $\rho_u/\rho_r \propto \log(y)^{-1/3}$ and does not approach a constant at early times, though at $T>\Lambda_{\mathcal{U}}$ the conformal symmetry is restored and one recovers $\rho_u=\sigma_{\mathcal{BZ}}T^4$. However, the dependence is so weak that $\rho_u/\rho_r \leq 0.022$ at the Planck scale , so it does significantly modify the BBN/CMB constraints. Therefore, the tracker solution is still effective in avoiding the fine tuning of the initial value.

Once we have applied all observational and theoretical constraints, we can discuss possible signatures of the model. Regarding the background observables, \eqref{eq:approxwu}, the biggest deviation from $\Lambda$CDM is in the $|\delta|\ll1$ regime. We find $-1\leq w_u\leq-0.985$ at best, at redshift $z\in[0,2]$. For example $y_0-1=10^{-4.5}, \delta=-0.1$ give $w_u\simeq-1+0.018z/(1+z)$. Obviously, the exponential approach to a CC at late times make it difficult to observe deviations from $\Lambda$CDM. One must either look for deviations at early times such as $\Delta N_{eff}\sim0.1$, cumulative effects over a large range of redshifts, or other observables that we turn to next.

\section{Perturbations in matter and unparticles} 
The essential smoking gun of any time-dependent DE model is its influence on the clustering in the Universe. We therefore calculate the growth of perturbations and compare the result to $\Lambda$CDM.
In a spatially flat universe, the evolution of the different density contrasts and gravitational potential is determined by: 
\begin{eqnarray}
 \Ddot{\tilde{\delta}}_i +  A_{i}  \dot{\tilde{\delta}}_{i} + B_{i}\tilde{\delta}_i &=& S_i \label{eq: muper1} \\
  \dot{\phi} + \left( 1 + \frac{k^{2}}{3 H^{2}}\right) \phi &=& -\frac{1}{2} \left( \Omega_m \tilde{\delta}_{m}  +  \Omega_{u} \tilde{\delta}_{u}\right),  \label{eq: muper2}
\end{eqnarray}  
where $i=u,m$ mark the density contrast of unparticles and matter respectively, $\phi$ is the gravitational potential,  and $\dot{}$ denotes differentiation with respect to e-folds. Since unparticles can still be expressed as $p(\rho)$, the adiabatic and effective speed of sound for unparticles are equal  $c_a^2\equiv\frac{\dot{p}}{\dot{\rho}}=\frac{\delta p}{\delta \rho}$ and for the matter component both speeds vanish. Hence, $A_{i} \,,\, B_{i}$ and $S_{i}$ are:
\begin{eqnarray}
 A_i &=& \frac{1}{2} \left[ 1 -3 w_{u}\, \Omega_{u} + 6\, c_{a_{i}}^{2} -12\, w_{i}\right] \nonumber \\ 
 B_{i} &=& \frac{3}{2}\left[ \left( c_{a_{i}}^{2} - w_{i}\right) \left( 1 - 3\, \Omega_{u} w_{u}  - 3\, w_{i}\right) + \frac{2 k^{2}}{3 H^{2}} c_{a_{i}}^{2} - 2\,\dot{w}_{i}\right] \nonumber \\
 S_{i} &=&  3 \left( 1 + w_{i} \right)  \Bigg[ \dot{\phi}\left( 1 + \frac{\dot{w}_{i}}{1 + w_{i}}\right) + \phi \left( 3\left( 1 + w_{u} \Omega_{u} \right) + \frac{2 k^{2}}{3 H^{2}} \right) \nonumber \\& +& \frac{3}{2}\left( \Omega_m \tilde{\delta}_{m}  +  \left( 1 +   c_{a_{i}}^{2} \right) \Omega_{u} \tilde{\delta}_{u}\right)\Bigg]
\end{eqnarray}
We solve the system of Eqs (\ref{eq: muper1}) and (\ref{eq: muper2}) using our background solution and the initial conditions of \cite{Mehrabi:2015hva}:
\begin{eqnarray}
 \tilde{\delta}_{m_{in}} &=& -2\, \phi_{i} \left( 1 + \frac{k^2}{3 H^{2}_{i}}\right)  \,,\nonumber \\
 \dot{\tilde{\delta}}_{m_{in}} &=&  - \frac{2}{3} \,\frac{k^{2}}{H_{i}^{2}} \,e^{N_{in}} \,\phi_{i}\,, \nonumber \\
 \tilde{\delta}_{u_{in}} &=&\left( 1 + w_{u_{in}}\right) \tilde{\delta}_{m_{in}} ,  \nonumber \\
 \dot{\tilde{\delta}}_{u_{in}} &=& \left( 1 + w_{u_{in}}\right) \dot{\tilde{\delta}}_{m_{in}} + \dot{w}_{u_{in}}\tilde{\delta}_{m_{in}}  .
\end{eqnarray}
\begin{figure}[]
\centering
\begin{adjustbox}{center}
\includegraphics[width=4.2cm, height=3.8cm]{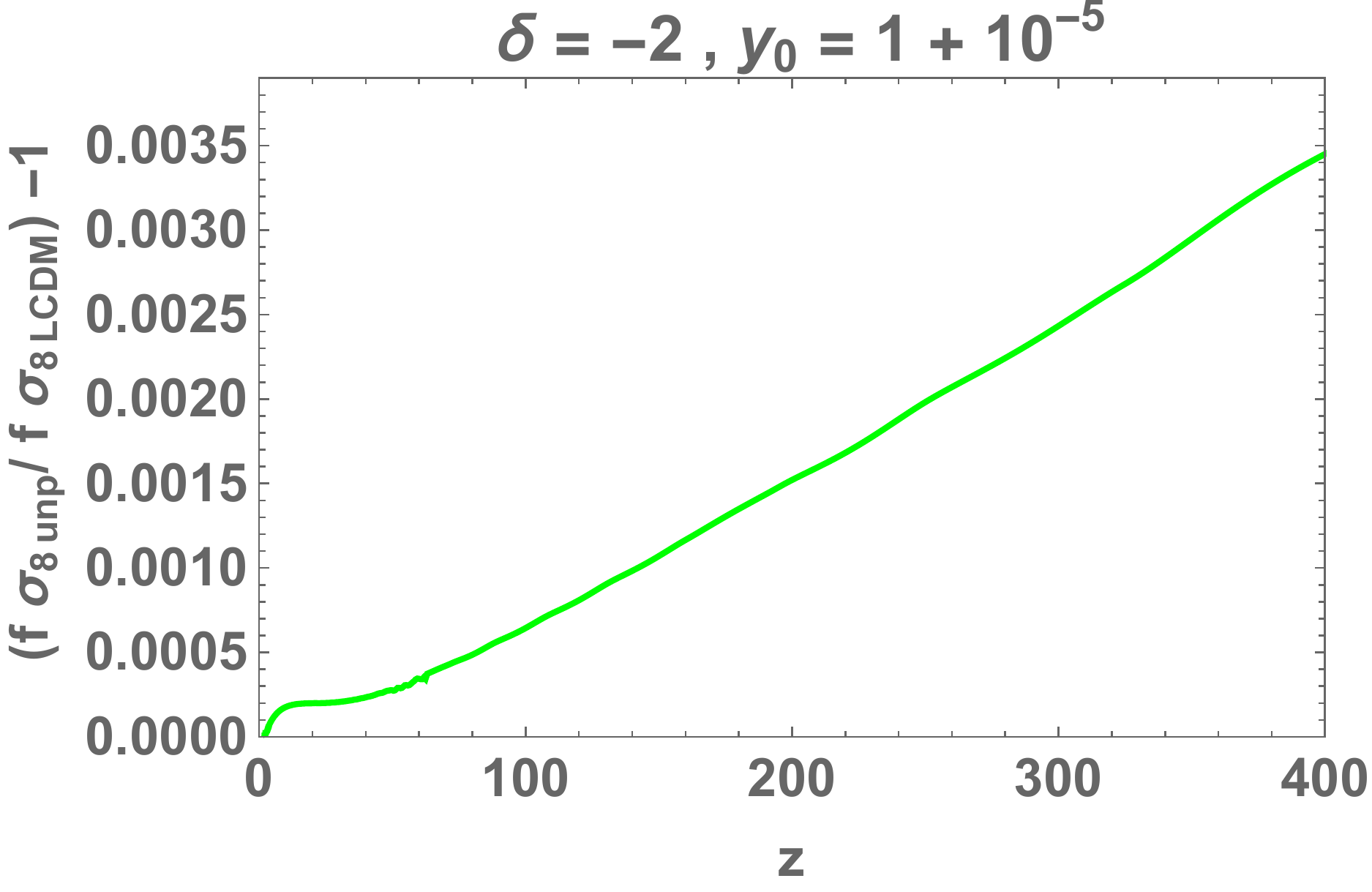} 
\includegraphics[width=4.2cm, height=3.8cm]{GA.pdf}
\end{adjustbox}
\caption{\it Left panel: Relative deviation of $f(z) \sigma_{8}$ of unparticles compared to $\Lambda$CDM for $0\leq z\leq400$. A significant deviation may be seen only at very large redshifts. Right panel: \it Comparison of the growth index, $\gamma$ as a function of redshift for our model compared to the $\Lambda$CDM prediction.}
\label{fig:pert1}
\end{figure}
 
Considering the linear growth of matter perturbations, $ D (z) = \frac{\tilde{\delta}_{m}(z)}{\tilde{\delta}_{m}(0)} $, one defines the growth rate of clustering $f\equiv\frac{d \log D}{d\log a}$ and the growth index $ \gamma(z) = \frac{d ln f(z)}{d \, \Omega_{m}}$. The left panel of Fig. \ref{fig:pert1} shows the relative difference of $ f(z)\, \sigma_{8} (z)$ between our model and $\Lambda$CDM, where $\sigma_{8}$ is the mass variance in a sphere of radius of $8 $ Mpc/h and can be written as $ \sigma_{8}(z) = \sigma_{8}(0) D(z) $, where $\sigma_{8}(z=0) $ is the present value from \cite{Aghanim:2018eyx}. The right panel of Fig. \ref{fig:pert1} shows the evolution of growth index for unparticles cosmology compared to the $\Lambda$CDM result. In all cases, the deviation from $\Lambda$CDM is at most $\sim 0.1\%$ at almost any given time, which is very difficult to detect unless there is an integrated effect. 
\section{Conclusions} \label{sec:conclusions}
We have investigated the evolution of a thermal average of a theory slightly shifted from its conformal fixed point.
In general, based on dimensional considerations, an anomalous dimension can result in a sector behaving as radiation at early times, and as a CC at late times, obtaining a DE model due to collective behavior. The model is technically natural, and has a built-in tracker mechanism. It can be defined at practically any energy scale $T_u<0.1 M_p$, and $10^{-30} M_p \leq B^{-1/\delta}\leq M_p$. Therefore, it solves the initial conditions and fine tuning problems. It avoids the no-dS conjecture, ameliorates the Hubble tension, and allows the true CC to vanish. We demonstrated the idea with a specific BZ/unparticles model.
The most severe tuning is the requirement of $\rho_u/\rho_r|_{BBN}\lesssim 0.1$ that predicts $y_0-1\lesssim 10^{-4.5}$ which is very mild. This additional $N_{eff.}$ are also the best chance of detection, absent some integrated effect. Two interesting future directions are a likelihood analysis of the current example and explicit calculations of the fundamental parameters $\sigma,B$ and $\delta$ in various CFTs, yielding much stronger predictions.

\end{document}